# Does light slowdown in dielectric media?

Vernon Cooray[1], Gerald Cooray[2], Farhad Rachidi[3], Marcos Rubinstein[4]

[1]Department of Electrical Engineering, Uppsala University, 752 37 Uppsala, Sweden.

[2]Karolinska Institute, Stockholm, Sweden

[3]Electromagnetic Compatibility Laboratory, Swiss Federal Institute of Technology (EPFL), 1015 Lausanne, Switzerland

[4]HEIG-VD, University of Applied Sciences and Arts Western Switzerland, 1401 Yverdon-les-Bains, Switzerland

## Abstract:

Observations and theoretical principles suggest that electromagnetic waves, including light, travel more slowly in dielectric media than in vacuum. Maxwell's equations, incorporating material-dependent permittivity and permeability, elegantly capture this effect. Previous studies indicate that the observed slower speed is due to interference effects, with the actual electromagnetic fields in the dielectric propagating at the speed of light in vacuum. However, these studies did not provide explicit expressions for the field components moving at this speed. The aim of the present study is to address this gap by analysing the structure of the electromagnetic field components within the dielectric. We examine how these components, each traveling at the speed of light in vacuum, interact to produce a net field that appears to propagate more slowly. Our findings show that the observed slower propagation in dielectric media results from interference between two types of waves: a forward-moving incident wave and a set of secondary waves, moving both in the forward (direction of propagation of the incident wave) and backward directions, and induced by the interaction of the incident wave with the dielectric medium. Both the incident wave and the secondary waves travel at the speed of light in vacuum. Importantly, we observe that the reflected wave caused by the impedance discontinuity at the boundary of the dielectric medium arises from secondary waves moving in the direction opposite to the incident wave.

## 1. Introduction

Maxwell's equations predict that the speed of electromagnetic wave propagation in dielectric media is slower than the speed of light in vacuum. The speed of electromagnetic waves in dielectric media is determined by the expression $v = 1/\sqrt{\mu_o \varepsilon_o \varepsilon_r}$, where $\varepsilon_r$ is the relative dielectric constant of the medium [1]. The relative dielectric constant relies on the intricate interaction of microscopic electromagnetic fields within the medium. Additionally, the macroscopic Maxwell's Equations have been formulated through statistical physical principles based on microscopic fields, as discussed in [2].

The fact that the actual speed of propagation of the waves in dielectric media is indeed equal to the speed of light in vacuum but the apparent slow speed of propagation is caused by the interaction of the waves with the ones induced by the dielectric material was pointed out previously by Feynman [3]. However, he did not provide a solution for the general case treated here. In [4], James and Griffith analyzed this problem by considering the secondary waves generated by the polarization of the dielectric medium due to the electric fields inside it. They began by examining the interaction between the incident field and the dielectric, calculating the resulting polarization current. From this current, they estimated the first-order secondary wave. This first-order secondary wave was then assumed to further polarize the dielectric, generating a second-order secondary wave. The second-order secondary wave, in turn, produced a third-order secondary wave, and so on. Through this iterative process, they summed up the secondary waves generated inside the dielectric by polarization currents. The resulting secondary waves were shown to constrain the incident wave to a front moving at a speed less than the speed of light. However, they did not give electromagnetic field expressions to identify the structure of the waves moving with the speed of light inside the dielectric. The goal of this paper is to extract the structure of these electromagnetic fields and to illustrate that the electromagnetic field inside the dielectric can be described by three waves moving with the speed of light in vacuum, one of which being the incident wave.

## 2. Methodology

As mathematically demonstrated by James and Griffin, the interaction between the incident wave, moving through the medium at the speed of light in vacuum, and the medium itself generates polarization currents. These currents produce the first-order secondary waves. The interaction of these first-order secondary waves subsequently generates additional polarization currents, resulting in second-order secondary waves. This process repeats ad infinitum. The cumulative effect of all secondary waves ultimately cancels the incident wave at all points beyond the front of the observed wave, which propagates through the medium at a speed less than that of light in vacuum.

In order to understand the structure of the electromagnetic waves propagating with the speed of light inside the medium, we start with the experimental observations, according to which, when an electromagnetic wave is incident on a dielectric medium, part of the wave will be reflected and the rest will be transmitted into the medium. The transmitted wave moves through the medium with a speed less that the speed of light in vacumm. In order to extract the structure of electromagnetic waves moving through the dielectric with the speed of light in vacuum, we have to consider the radiation from the time varying polarization currents in the medium. The scenario we will consider for this procedure is the following. The incident electromagnetic wave passes through the medium with a speed equal to that of light in vacuum and its interaction with the dielectric will give rise to polarization currents which will radiate secondary waves in the medium. These secondary waves, again moving with the speed of light in vacuum will interfere with the incident wave to produce a net wave that moves through the medium with a speed less than the speed of light in vacuum.

To analyse this further, the dielectric medium is divided into infinitesimally thin sheets located parallel to the dielectric boundary. From experimental observation, we can extract the polarization currents generated in each of these infinitesimal sheets of dielectric and, from them, we will be able to determine the magnitude and phase of the secondary waves. The polarization current from a given thin sheet will radiate both in the forward and backward directions giving rise to secondary waves moving with the speed of light in vacuum. By leveraging the fact that the forward-moving secondary waves cancel the incident wave at all points beyond the wave front of the observed wave (i.e., the wave moving with speed $v$), we will be able to obtain the relationship between the speed $v$ and the dielectric constant of the dielectric that is necessary to achieve this nullification.

First, we will consider an electromagnetic wave incident normal to the boundary and, based on the results, we will extend the results to oblique incidence.

## 3. Normal incidence

### 3.1 The polarization currents induced inside the dielectric medium

Let us denote the incident electric field at the boundary of the dielectric medium as $E_{inc} = E_0 \sin \omega(t - z / c)$. The dielectric medium is assumed to be semi-infinite, with its boundary coinciding with the x-y plane at z = 0. Part of this field will be reflected and the rest of the wave will be transmitted through the dielectric medium. From Maxwell's equations confirmed by experimental observations, we know that the wave transmitted through the medium, that is the wave observed inside the dielectric, is given by

$$E_{obs} = E_0 \frac{2}{1+\sqrt{\varepsilon_r}} \sin\omega\{t - z/v\} \qquad (1)$$

In the above equation $\varepsilon_r$ is the relative dielectric constant of the medium. Our goal is to find out how this wave is generated by electromagnetic fields moving with the speed of light in vacuum. Consider an infinitesimally thin dielectric sheet parallel to the boundary at a distance $\xi$ from it. The polarization current density induced by this electric field in the dielectric sheet is given by (taking the derivative of the polarization vector) [1, 4]

$$dJ_{pol}(\xi,t) = \varepsilon_0(\varepsilon_r - 1)E_0\omega \frac{2}{1+\sqrt{\varepsilon_r}} \cos\omega\{t - \xi/v\}\, d\xi \qquad (2)$$

This time-varying polarization current radiates secondary waves into the dielectric medium that propagate both in the forward direction (deeper into the dielectric) and in the backward direction (toward the dielectric boundary). Treating the infinitesimally thin dielectric layer as a thin current sheet, the source field for the forward-moving secondary wave generated by this polarization current can be expressed as [4]

$$dE_{sf}(\xi,t) = -\frac{\mu_0 c}{2}\varepsilon_0(\varepsilon_r - 1)E_0\omega \frac{2}{1+\sqrt{\varepsilon_r}} \cos\omega\{t - \xi/v\}\, d\xi \qquad (3)$$

This can be further simlified to obtain

$$dE_{sf}(\xi,t) = -\frac{1}{c}\left(\sqrt{\varepsilon_r} - 1\right)\omega E_0 \cos\omega\{t - \xi/v\}\, d\xi \qquad (4)$$

Now, the incident wave travels through the dielectric medium with the speed of light in vacuum whereas the observed wave travels through the medium at speed $v$. In order to satisfy this observation, the forward moving secondary waves (i.e., moving in the direction of the incident wave) should nullify the incident wave at all points beyond the wavefront of the observed wave. Consider a given time I. At this moment, the front of the observed wave is located at a distance $z_f = vt$ from the boundary of the dielectric medium. Recall that each infinitesimal layer will contribute to both forward-moving and backward-moving secondary waves. The forward-moving secondary wave at location $z_f$ can be obtained by summing the contributions of all source fields located within $0 \le z \le z_f$, using Equation (4). The result is

$$E_{sf}(z_f,t) = -\frac{1}{c}\left(\sqrt{\varepsilon_r} - 1\right)\omega E_0 \int_0^{z_f} \cos\omega\{t - \xi/v - (z_f - \xi)/c\}\, d\xi \qquad (5)$$

After performing the integral we obtain

$$E_{sf}(z_f,t) = \frac{1}{c}\left(\sqrt{\varepsilon_r} - 1\right)\frac{1}{\left(\frac{1}{v} - \frac{1}{c}\right)} E_0 \sin\omega\{t - z_f/c\} \qquad (6)$$

Now, in order for the forward moving secondary wave to nullify the incident wave at all points at or beyond wave front of the observed wave, the forward moving secondary wave should satisfy the condition

$$\frac{1}{c}\left(\sqrt{\varepsilon_r}-1\right)\frac{1}{\left(\frac{1}{v}-\frac{1}{c}\right)}E_0\sin\omega\left\{t-z_f/c\right\}=E_0\sin\omega\left\{t-z_f/c\right\} \quad (7)$$

This condition can be satisfied only if the speed of propagation of the observed wave is connected to the dielectric constant of the medium through the relationship

$$v=\frac{c}{\sqrt{\varepsilon_r}}=\frac{1}{\sqrt{\varepsilon_0\mu_0\varepsilon_r}} \quad (8)$$

This is exactly what we would have obtained had we used the Maxwell's equations directly to solve the problem. Note that since both the incident and the forward-moving secondary waves are moving with the speed of light in vacuum, if they nullify each other at the front of the observed wave, they will also nullify each other at all points ahead of the front of the observed wave.

Now we are in a position to explicitly derive the expressions for the field components moving at the speed of light in vacuum within the dielectric. The three field components are, namely, the forward-moving secondary wave, the backward-moving secondary wave and the forward-moving incident wave.

### 3.2 The forward and backward-moving secondary waves

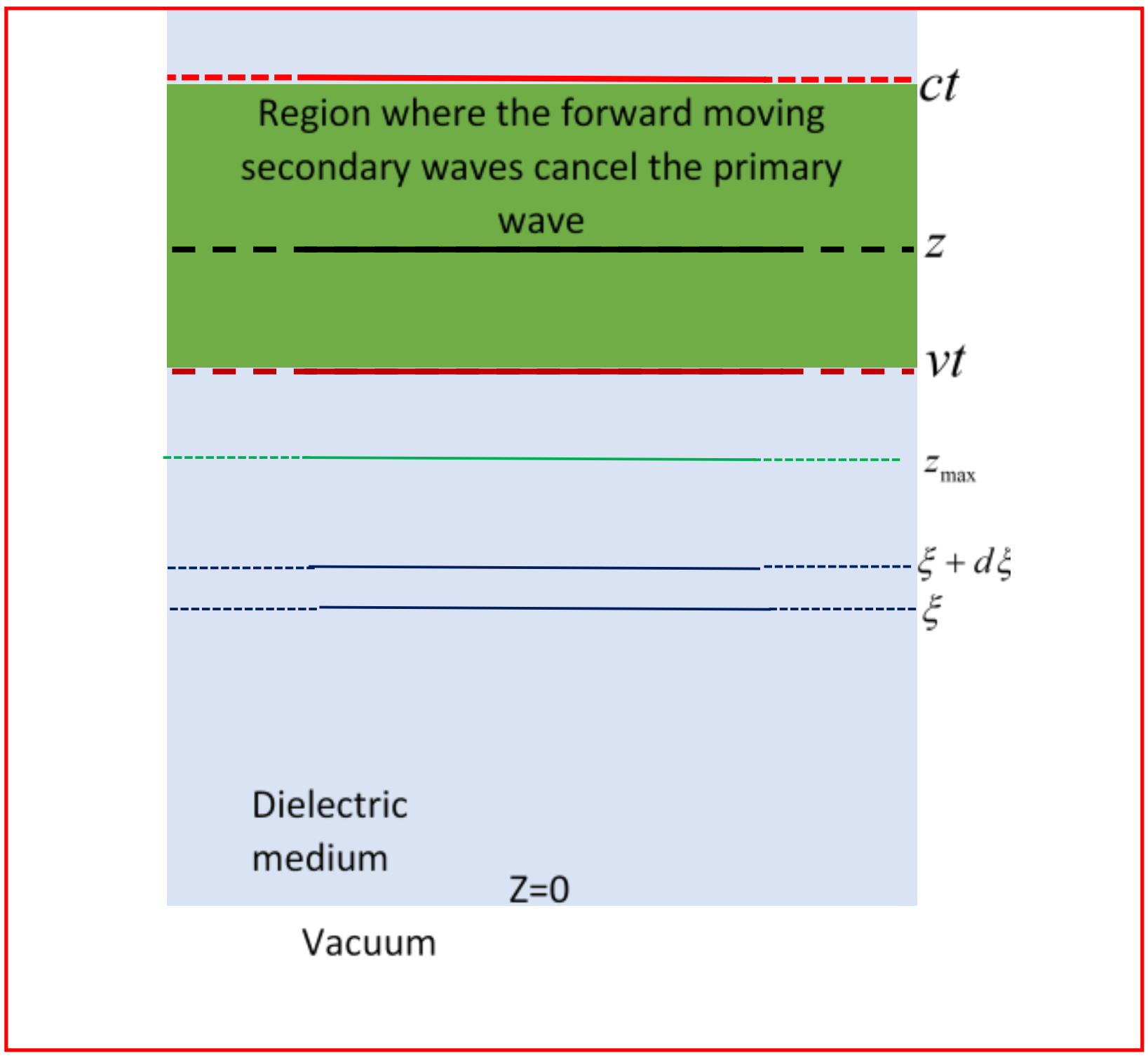

Figure 1: Geometry pertinent to the evaluation of the forward-moving secondary wave at any location. The situation shown in the diagram corresponds to time $t$ assuming that the incident wave has entered the dielectric medium (i.e., $z$ = 0) at time $t=0$. At this time, the observed wave has propagated a distance $vt$ into the dielectric medium and the incident wave, which is propagating at the speed of light in vacuum has propagated a distance $ct$ into the dielectric medium. The region where the forward-moving secondary wave cancels the incident wave is marked in green colour. Note that level $z$ is located such that $vt < z \le ct$.

Let us first examine the forward-moving secondary wave. Consider the events taking place at time $t$. The situation is depicted in Figure 1. The incident wave has propagated a distance $ct$ across the medium and the forward-moving secondary wave has also travelled the same distance. The net wave or the observed wave has travelled a distance $vt$. First, we will estimate the amplitude of the forward-moving secondary wave at a level $z$ where $z > vt$. The forward-moving secondary wave at this level is generated by the sum of all elementary forward-moving secondary waves generated along the path of the electromagnetic wave (Equation (4)). The total contribution of all infinitesimal layers to the electric field at level $z$ and at a time $t$ is given by (note that we have replaced $\sqrt{\varepsilon_r}$ by $c/v$)

$$E_{sf}(z,t) = -\int_0^{z_{\max}} E_0\omega\left(\frac{1}{v}-\frac{1}{c}\right)\cos\left\{\omega\left(t-\xi/v-\frac{(z-\xi)}{c}\right)\right\}d\xi \quad (9)$$

In the above integral $z_{\max}$ is the limiting value of $\xi$ that can make a contribution to the secondary wave at $z$. Note that at time $t$, elements located beyond $z_{\max}$ do not have enough time for their effect to be felt at level $z$. The value of $z_{\max}$ is given by

$$z_{\max} = (t - z/c)/\left(1/v - 1/c\right) \quad (10)$$

After performing the integral, we obtain

$$E_{sf}(z,t) = -E_0\sin\{\omega(t-z/c)\} \qquad z \ge vt \quad (11)$$

As described in the previous section, observe that this wave will cancel out the incident wave at all points located beyond the front of the observed wave.

We will now estimate the secondary wave below the front of the observed wave. To that aim, the integral has to be changed as follows

$$E_{sf}(z,t) = -\int_0^{z} E_0\omega\left(\frac{1}{v}-\frac{1}{c}\right)\cos\left\{\omega\left(t-\xi/v-\frac{(z-\xi)}{c}\right)\right\}d\xi \qquad z < vt \quad (12)$$

The solution of which is

$$E_{sf}(z,t) = -E_0\left[\sin\{\omega(t-z/c)\} - \sin\{\omega(t-z/v)\}\right] \qquad z < vt \qquad (13)$$

Let us now derive an expression for the backward-moving secondary wave. As before, consider the events taking place at time $t$. The situation for this derivation is illustrated in Figure 2. The incident wave has propagated a distance $ct$ across the medium and the forward-moving secondary wave has also travelled the same distance. The observed wave has travelled a distance $vt$. The level where we would like to estimate the amplitude of the backward-moving secondary wave is $z$. The backward-moving secondary wave at this point is generated by the sum of all elementary secondary waves generated at all points located beyond the level $z$. Consider an element $d\xi$ located at height $\xi$. The contribution to the backward-moving secondary wave at $z$ produced by this element is

$$dE_{sb}(z,t) = -\omega E_0\left(\frac{1}{v}-\frac{1}{c}\right)\cos\left\{\omega\left(t-\xi/v-\frac{(\xi-z)}{c}\right)\right\}d\xi \quad (14)$$

The total contribution at $z$ is given by

$$E_{sb}(z,t) = -\int_{z}^{z_{\max}} E_0 \omega \left(\frac{1}{v} - \frac{1}{c}\right) \cos\left\{\omega\left(t - \xi / v - \frac{(\xi - z)}{c}\right)\right\} d\xi \quad (15)$$

It is important to note that any dielectric layer situated beyond $z_{\max}$ does not play a role in generating the backward-moving secondary wave at $z$. The value of $z_{\max}$ is given by

$$z_{\max} = (t + z / c) / \left(1 / v + 1 / c\right) \quad (16)$$

After performing the integral in Equation (15), we obtain

$$E_{sb}(z,t) = -\frac{\left(\frac{1}{v} - \frac{1}{c}\right)}{\left(\frac{1}{v} + \frac{1}{c}\right)} E_0 \sin\left\{\omega(t - z / v)\right\} \quad (17)$$

Note that this equation mimics a wave moving in the forward direction. However, it consists of a multitude of secondary waves all moving in a backward direction with the speed of light in vacuum.

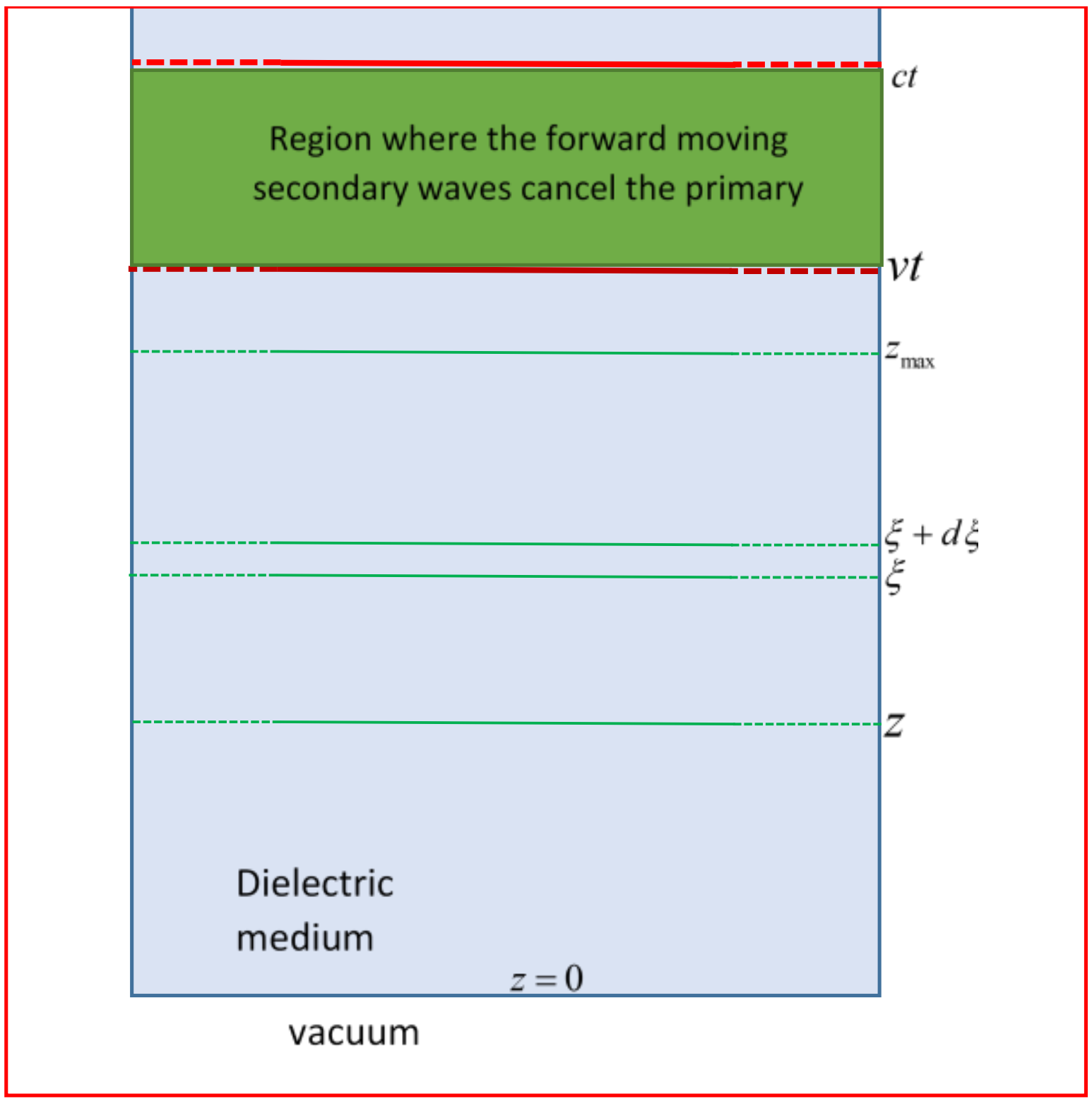

Figure 2: Geometry pertinent to the evaluation of the backward-moving secondary wave at any location. The situation is identical to that of Figure 1 except that level $z$ is located such that $z < vt$.

### 3.3 The structure of the electromagnetic waves inside the dielectric medium

According to the analysis presented in the previous sections, the electromagnetic wave structure inside the dielectric for normal incidence can be described as a sum of three waves, all moving with the speed of light in vacuum. The three waves are the incident wave, the forward-moving secondary wave, and the backward-moving secondary wave. The incident wave is given by

$$E_{inc}(z,t) = E_0 \sin\{\omega(t - z/c)\} \qquad (18)$$

The forward-moving secondary wave is given by

$$E_{sf}(z,t) = -E_0 \sin\{\omega(t - z/c)\} \qquad z \geq vt \quad (19)$$

$$E_{sf}(z,t) = E_0 \sin\{\omega(t - z/v)\} - E_0 \sin\{\omega(t - z/c)\} \quad z < vt \qquad (20)$$

And the backward-moving secondary wave is given by

$$E_{sb}(z,t) = -\frac{\left(\frac{1}{v} - \frac{1}{c}\right)}{\left(\frac{1}{v} + \frac{1}{c}\right)} E_0 \sin\{\omega(t - z/v)\} \qquad z < vt \qquad (21)$$

The sum of the three waves gives rise to the observed wave inside the dielectric and it is given by

$$E_{obs}(z,t) = \left[1 - \frac{\left(\frac{1}{v} - \frac{1}{c}\right)}{\left(\frac{1}{v} + \frac{1}{c}\right)}\right] E_0 \sin\{\omega(t - z/v)\} \qquad (22)$$

Since the forward-moving secondary wave interferes destructively with the incident wave at all points beyond the front of the observed wave, no electromagnetic wave exists above the front of the observed wave.

### 3.4 The reflected wave at the boundary of the dielectric

Let us consider the backward-moving secondary wave at locations below the boundary, i.e., $z < 0$. Consider any arbitrary level $z$ where $z < 0$. The backward-moving wave at this location (i.e., the reflected wave) is given by

$$E_{ref}(z,t) = -\int_0^{z_{\max}} E_0 \omega \left(\frac{1}{v} - \frac{1}{c}\right) \sin\left\{\omega\left(t - \xi/v - \frac{(\xi - z)}{c}\right)\right\} d\xi \quad z < 0 \qquad (23)$$

The solution of which is

$$E_{ref}(z,t) = -\frac{\left(\frac{1}{v} - \frac{1}{c}\right)}{\left(\frac{1}{v} + \frac{1}{c}\right)} E_0 \sin\{\omega(t + z/c)\} \qquad z < 0 \qquad (24)$$

This waveform signifies a wave emanating from the dielectric medium and advancing at the speed of light in vacuum in a direction opposite to that of the incident wave. As soon as the incident wave reaches the dielectric boundary, this wave emerges and extends beyond the dielectric medium. It can be recognized as the reflected wave induced by the impact of the incident wave at the dielectric medium's boundary. Therefore, the analytical framework presented here accommodates the reflected wave generated at the boundary of the dielectric medium.

## 4. Oblique incidence

When the angle of incidence deviates from normal to the dielectric boundary, two scenarios must be considered: TE incidence, where the electric field is perpendicular to the plane of incidence, or TM incidence, where the magnetic field is perpendicular to the plane of incidence. In TE incidence, the electric field remains parallel to the dielectric boundary, allowing for an analysis similar to that conducted previously to extract the relevant field components. Correspondingly, the field components relevant to the TM mode can be obtained using the principle of duality [1].

The pertinent geometry for the case of TE incidence is illustrated in Figure 3. Given that the boundary coincides with x-y plane, the component of the wave of interest to us propagates along the z-direction. The phase velocity of the portion of the wave moving along the z-axis is given by $c/\cos\theta_i$ for the incident wave in vacuum and $v/\cos\theta_r$ for the observed wave inside the dielectric. Let us denote these speeds by $\tilde{c}$ and $\tilde{v}$. Let us consider the incident wave. It is denoted by (here the electric field is directed along the negative y-axis)

$$E_{inc} = E_0 \sin\left\{\omega(t - x\sin\theta_i / c - z\cos\theta_i / c)\right\} \quad (25)$$

The motion of the wave along the z-direction is given by

$$E_{inc} = E_0 \sin\left\{\omega(t - z/\tilde{c})\right\} \quad (26)$$

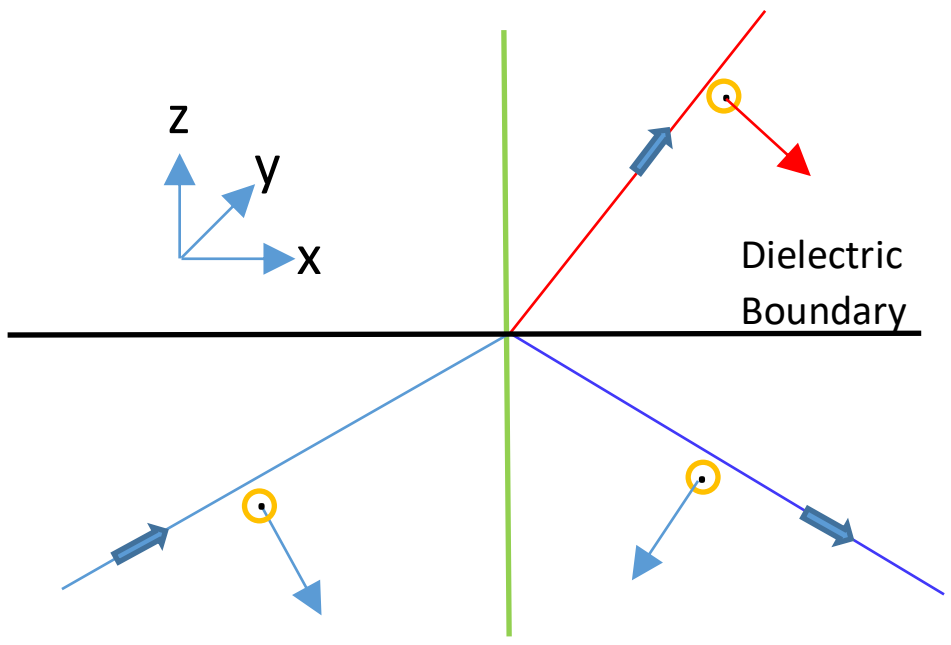

Figure 3. The geometry relevant to the oblique incidence of a TE electromagnetic wave on a dielectric boundary. The electric fields of the waves are directed along the negative *y* axis, which is directed perpendicular to the plane of incidence.

Thus, once we know the expression for the portion of the wave moving along the z-direction, the full wave can be reconstructed easily by using the relationship between them as given by equations (25) and (26). Now, the theory developed for normal incidence where the wave propagates along the *z*-axis can be directly applied to the wave defined by Equation (26). Using the previous notation, the forward $(E_{sf-z})$ and backward $(E_{sb-z})$ moving secondary waves are described by

$$E_{sf-z}(z,t) = -E_0 \sin\left\{\omega(t - z/\tilde{c})\right\} \quad z > \tilde{v}t \quad (27)$$

$$E_{sf-z}(z,t) = -E_0\left[\sin\left\{\omega(t - z/\tilde{c})\right\} - \sin\left\{\omega(t - z/\tilde{v})\right\}\right] \quad z \le \tilde{v}t \quad (28)$$

$$E_{sb-z}(z,t) = -\frac{\left(\frac{1}{\tilde{v}} - \frac{1}{\tilde{c}}\right)}{\left(\frac{1}{\tilde{v}} + \frac{1}{\tilde{c}}\right)} E_0 \sin\{\omega(t - z/\tilde{v})\} \quad (29)$$

The subscript $z$ in the above Equation denotes that we are considering the movement of the wave along the $z$-direction. The net wave or the observed wave inside the dielectric is given by the sum of the three waves and it is given by

$$E_{obs-z}(z,t) = \left[1 - \frac{\left(\frac{1}{\tilde{v}} - \frac{1}{\tilde{c}}\right)}{\left(\frac{1}{\tilde{v}} + \frac{1}{\tilde{c}}\right)}\right] E_0 \sin\{\omega(t - z/\tilde{v})\} \quad (30)$$

The backward moving secondary wave in the region $z > 0$ is given by

$$E_{sb-z}(z,t) = -\frac{\left(\frac{1}{\tilde{v}} - \frac{1}{\tilde{c}}\right)}{\left(\frac{1}{\tilde{v}} + \frac{1}{\tilde{c}}\right)} E_0 \sin\{\omega(t + z/\tilde{c})\} \quad z < 0 \quad (31)$$

The field expressions given above are for the $z$-component of the wave. Using the correspondence bewteen the z-component and the full wave as given by equations (25) and (26), the full expressions for the waves moving inside the medium and the reflected wave can be writen as:

$$E_{inc-TE}(z,t) = E_0 \sin\{\omega(t - z\cos\theta_i/c - x\sin\theta_i/c)\} \quad (32)$$

$$E_{sf-TE}(z,t) = -E_0 \sin\{\omega(t - z\cos\theta_i/c - x\sin\theta_i/c)\} \quad z \geq \tilde{v}t \quad (33)$$

$$E_{sf-TE}(z,t) = -E_0\left[\sin\{\omega(t - z\cos\theta_i/c - x\sin\theta_i/c)\} - \sin\{\omega(t - z\cos\theta_r/v - x\sin\theta_r/v)\}\right] \quad z < \tilde{v}t \quad (34)$$

$$E_{sb-TE}(z,t) = -\frac{\left(\frac{\cos\theta_r}{v} - \frac{\cos\theta_i}{c}\right)}{\left(\frac{\cos\theta_r}{v} + \frac{\cos\theta_i}{c}\right)} E_0 \sin\{\omega(t - z\cos\theta_r/v - x\sin\theta_r/v)\} \quad (35)$$

Equations (32) to (35) describe the incident wave, the forward moving secondary wave, and the backward moving secondary wave, respectively. The observed wave inside the dielectric and the reflected wave are given, respectively, by

$$E_{obs-TE}(z,t) = \left[1 - \frac{\left(\frac{\cos\theta_r}{v} - \frac{\cos\theta_i}{c}\right)}{\left(\frac{\cos\theta_r}{v} + \frac{\cos\theta_i}{c}\right)}\right] E_0 \sin\{\omega(t - z\cos\theta_r/v - x\sin\theta_r/v)\} \quad (37)$$

$$E_{ref-TE}(z,t) = -\frac{\left(\frac{\cos\theta_r}{v} - \frac{\cos\theta_i}{c}\right)}{\left(\frac{\cos\theta_r}{v} + \frac{\cos\theta_i}{c}\right)} E_0 \sin\{\omega(t + z\cos\theta_i/c - x\sin\theta_i/c)\} \quad (38)$$

These field components are in perfect agreement with the predictions of classical electrodynamics for TE incidence.

## 5. Discussion

The analysis presented herein reveals that the propagation of an electromagnetic wave or a light beam through a dielectric medium can be conceptualized as the interaction of three waves moving at the speed of light in vacuum. These waves include the incident wave, along with two secondary waves generated through the interaction of the electromagnetic wave's electric field with the dielectric medium. Among these secondary waves, one travels backward in relation to the incident wave, while the other travels in the same direction as the incident wave. The secondary wave moving backward corresponds to the reflected wave when the incident wave encounters the boundary of the dielectric medium. Importantly, the incident and forward-moving secondary waves exhibit mutual cancellation at all points beyond the wavefront of the observed wave. It is noteworthy that while Equation (17) characterizes a wave in motion at a speed $v$, as elucidated by Equation (15), this wave comprises a composite of elementary backward-moving secondary waves, each propagating at the speed of light in vacuum. It is essential to recognize that these elementary waves originate at the forefront of the observed wave, which itself is advancing at a speed $v$.

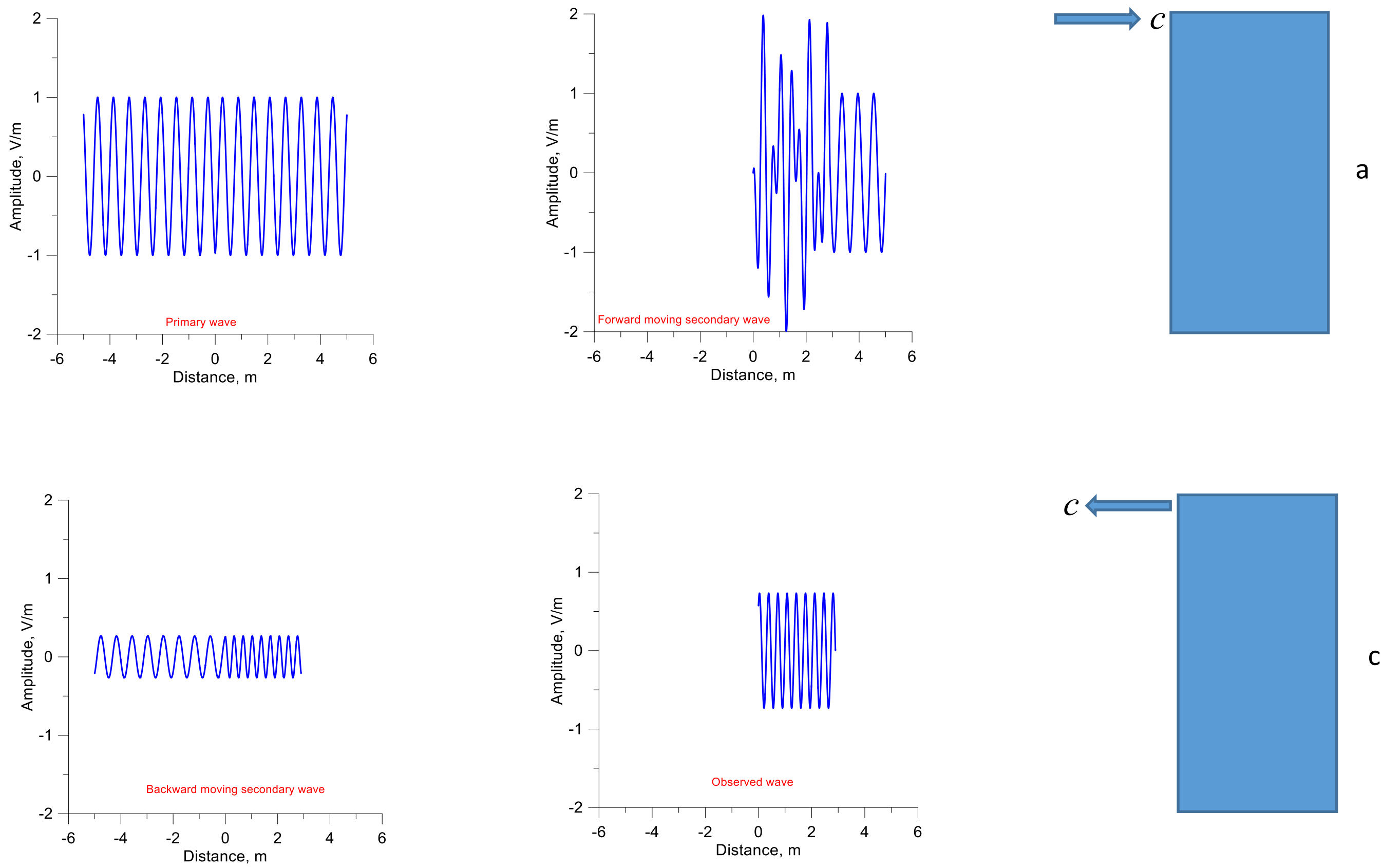

Figure 4: Different components of the electromagnetic waves inside and outside the dielectric medium for normal incidence at one particular moment in time. The shaded area represents the dielectric medium. In constructing this diagram, the frequency of the wave was taken to be $5\times10^8$ Hz and the relative dielectric constant was fixed at 3. In the illustration the time was fixed at 5/c seconds with $t = 0$ corresponding to the time when the wave is incident on the dielectric boundary. (a) Incident wave. (b) Forward-moving secondary wave. (c) Backward-moving secondary wave. (d) Sum of the field components shown in (a), (b) and (c) inside the dielectric.

The field components pertinent to normal incidence at a given instant of time are illustrated in Figure 4. The waveforms shown in this figure are derived from equations (18), (19), (20), and (21), which represent the forward-moving secondary wave, the backward-moving secondary wave, and the incident wave, respectively. These equations describe both the spatial and temporal variations of the electric field associated with these waves. To visualize these waves, we selected the frequency of these waves to be $5\times10^8$ Hz and fixed the relative dielectric constant of the medium at 3. For illustrating the spatial variation of the waves at a given moment, we set the time in equations (18), (19), (20), and (21) arbitrarily to $5.0/c$ seconds, where $c$ is the speed of light in vacuum. Since we assume that time $t=0$ corresponds to the time when the wave is incident on the dielectric boundary, the incident wave would have propagated 5 meters into the dielectric medium by this time.

With the time fixed, these equations describe how the amplitude of the electric field of the waves varies as a function of space (i.e., along the z-axis). These amplitudes are depicted in Figure 4. Figure 4a shows how the amplitude of the incident wave changes with distance. Figure 4b illustrates the amplitude of the forward-moving secondary wave. Figure 4c displays the amplitude of the backward-moving secondary wave. Figure 4d presents the sum of the incident wave, the forward-moving secondary wave, and the backward-moving secondary wave inside the dielectric medium. The resulting wave inside the dielectric medium is identical to the observed wave as given by Equation (22).

If the E-field is polarized along the *x*-axis, the component of the B-field directed along the *y*-axis can be obtained from $\partial B_y/\partial t = -\partial E_x/\partial z$. These field components are in exact agreement with the predictions of classical electromagnetics. However, all the components of the wave are moving with the speed of light in vacuum.

In the scenario of oblique incidence, it is notable that the incident wave persists in its propagation through the medium along the same direction at which it initially struck the dielectric boundary. Additionally, it is evident that the wave fronts of the secondary waves moving both forward and backward are inclined at an angle to the *z*-axis. This inclination can be comprehended by considering that the phase of the electric field, which triggers the secondary waves from a particular dielectric layer, undergoes variation along the *x*-axis at any given moment. It is this phase alteration that leads to the emergence of secondary waves with wave fronts inclined to the *z*-axis. For example, the direction of propagation of the forward-moving secondary wave generated by any dielectric layer makes an angle with the *z*-axis, the tangent of which is given by $\frac{\sin\theta_r}{v}/\frac{\cos\theta_i}{c}$. However, since $\frac{\sin\theta_r}{v}=\frac{\sin\theta_i}{c}$, the forward-moving secondary wave makes an angle $\theta_i$ with the z-axis.

Once Maxwell's equations are provided, numerous paths can lead to the same solution in problems of electrodynamics [3]. This is true for the problem we have analyzed as well. Given Maxwell's equations, one can conclude that the observed speed of light in dielectric media is less than the speed of light in vacuum through various approaches. Although the final observation remains the same, the path to that conclusion is not unique. In this paper, we explore an alternative path that leads to the same solution. Our selected path demonstrates that the final result can also be explained by expressing the total electromagnetic wave as a sum of three components, which propagate with the speed of light in vacuum and interfere with each other.

While one might argue that our work is of academic interest only, since standard procedures also reach the same final result, exploring different paths can provide deeper insights into the physical phenomena occurring during the interaction of electromagnetic waves with dielectric media. A similar situation arises in the calculation of electromagnetic fields from given charge and current distributions. Multiple methods can yield the same solution, each illuminating the richness of Maxwell's theory. For instance, one can use either dipole approximation or field equations related to the movement and acceleration of charges to arrive at the same conclusion [5]. However, in specific applications, either the dipole approximation or the charge acceleration method might be preferred.

Thus, exploring various physical scenarios that lead to the same conclusion can be beneficial for different applications.

One example where our outlined procedure is useful is in understanding the mechanism of lightning flashes. Lightning flashes are driven by electromagnetic waves moving at the speed of light in vacuum. However, these waves interact along the lightning channel to generate a wavefront that propagates slower than the speed of light. Investigating this mechanism using standard procedures is challenging, but our method can easily explain the reduced speed of propagation [6]. Moreover, this perspective provides valuable insights into the interaction of these electromagnetic waves with the hot lightning channel, which is valuable for modeling lightning flashes.

In our analysis, we considered a semi-infinite, ideal dielectric. It would be of interest to study the behavior of the three waves as they pass through a dielectric slab. Additionally, exploring how these waves are modified inside a lossy dielectric is also of interest. Moreover, our analysis is for an incident wave which varies sinusoidally. We are in the process of investigating the structure of the electromagnetic fields inside the dielectric when the wave incident on the boundary is a transient.

## 6. Conclusion

This study builds on prior research demonstrating that the apparent reduction in electromagnetic wave speed within dielectric media is due to interference effects, rather than an actual decrease in propagation speed. Here, we reveal the detailed structure of waves propagating at the speed of light within a dielectric medium. Our results indicate that three distinct waves travel at the speed of light in vacuum inside the dielectric: a forward-moving secondary wave, a backward-moving secondary wave, and the incident wave itself. The forward-moving secondary wave cancels the forward-moving component of the incident wave above the advancing observed wavefront, which travels at a speed lower than the speed of light in vacuum. This interference effectively limits the observed speed of the resultant wave. Our study shows that for this cancellation to happen, the speed of propagation of the observed wave, $v$, should be related to the relative dielectric constant of the medium. The derived connection is in agreement with standard theory. Additionally, our study clarifies that the backward-moving portion of the secondary wave corresponds to the reflected wave, generated by the impedance mismatch at the boundary where the incident wave enters the dielectric medium.